\documentclass[twocolumn,aps,prb,epsf]{revtex4}
\usepackage{graphicx}
\begin{document}

\title{Self-consistent renormalization theory of spin fluctuations in paramagnetic spinel  LiV$_2$O$_4$}

\author{V. Yushankhai$^{1,2}$, P. Thalmeier$^{1}$, and  T. Takimoto $^{1}$}

\affiliation{ $^1$Max-Planck-Institut f\"ur Chemische Physik fester Stoffe, D-01187 Dresden,
  Germany\\ $^2$Joint Institute for Nuclear Research, 141980 Dubna, Russia}

\date{\today}

%%%%%%%%%%%%%%%%%---  Abstract ---%%%%%%%%%%%%%%%%%%

\begin{abstract}
A phenomenological description for the dynamical spin susceptibility   $\chi\left({\bf q},\omega;T\right)$
observed in inelastic neutron scattering  measurements on powder samples of  LiV$_2$O$_4$
is developed in terms of the parametrized self-consistent renormalization (SCR) theory of spin fluctuations.
Compatible with previous studies at $T\to 0$, a peculiar distribution in  ${\bf
q}$-space of strongly enhanced and slow spin fluctuations at $q \sim Q_c
\simeq$ 0.6 $\AA^{-1}$ in LiV$_2$O$_4$   is involved to derive the mode-mode coupling term
entering the basic equation of the SCR theory. The equation is solved self-consistently  with the
parameter values found from a fit of  theoretical results to experimental data.
For low temperatures,  $T \lesssim 30$K, where the
SCR theory is more reliable,  the  observed temperature
variations of the static spin susceptibility $\chi\left(Q_c;T\right)$ and the relaxation rate
$\Gamma_Q\left(T\right)$ at $q\sim Q_c$ are 
well reproduced by those suggested by the theory. For  $T\gtrsim 30$K, the present SCR is capable 
in predicting only main trends in $T$-dependences of  $\chi\left(Q_c;T\right)$ and $\Gamma_Q\left(T\right)$.
 The  discussion is focused on a  marked evolution  (from $q \sim Q_c$ at $T\to 0$ towards low $q$ values  at higher temperatures)   of the dominant low-$\omega$ integrated 
neutron scattering intensity $I\left( q; T\right)$.
\end{abstract}

\pacs{71.27.+a, 71.10.-w,  71.10.Fd,  71.30.+h}

\maketitle

%%%%%%%%%%%%%%%%--- I ----%%%%%%%%%%%%%%%%%%%%%%%%%%%

\section{Introduction}
The metallic vanadium oxide   LiV$_2$O$_4$ was found~ \cite{Kondo97,Johnston00,Kondo99}  to be the first example of 3$d$-electron system with heavy fermion (HF) behavior.   LiV$_2$O$_4$ has the cubic spinel structure with the magnetic vanadium ions (in the mixed valence state V$^{3.5+}$) occupying the pyrochlore lattice sites. Mechanisms for formation of heavy quasiparticles in this
strongly correlated electronic system are still under debate~\cite{Fulde03,Arita07}.  The geometrical frustration of the pyrochlore lattice is likely to be a crucial  aspect of the problem. \cite{Lacroix01,Fulde01,Burdin02,Hopkinson02,Fujimoto02,Yamashita03,Laad03}
The frustration can be directly related to the suppression of a long-range magnetic order at any $T$ and instead, the system  is close to a magnetic instability. Thus, the enhanced low-energy dynamic spin fluctuations are expected to influence considerably the low-$T$ properties of  LiV$_2$O$_4$, leading to formation of HF behavior. 
 
 Low-$\omega$ spin fluctuations in  LiV$_2$O$_4$ were studied in a series of inelastic neutron scattering (INS)
 measurements,~\cite{Krimmel99,Lee01,Murani04}  as well as in NMR experiments~\cite{Fujiwara98,Mahajan98,Fujiwara04,Johnston05} 
  in a wide range of temperatures 
from 0.5 to $\sim 700$K. In INS measurements  on polycrystalline
 samples of LiV$_2$O$_4$ at low temperatures, $T\lesssim$ 2K,
 short-range antiferromagnetic (AFM) correlations with a characteristic relaxation rate $\hbar\Gamma\sim $1meV 
 in a broad region of  wavevectors around  $q \sim Q_c \simeq$ 0.6 $\AA^{-1}$ were
 observed. As the temperature
 is increased above 2K (to $\sim60$K and higher),  the integrated  scattering
 intensity at low energy transfer $\hbar \omega
 \lesssim$1meV was found to be shifted toward low ${\bf q}$ values.  This indicates that short-range AFM  correlations get suppressed in favor of those located near the Brillouin zone (BZ) center .
 
 At a momentum transfer ${\bf q}$, the measured integrated
 intensity is defined as
 %01
\begin{equation}
I\left({\bf q}; T\right)=\int_{\omega_1}^{\omega_2} S\left({\bf q}, \omega; T\right). \label{a1}
\end{equation}
Here, $\hbar \omega_1 \simeq$0.2meV is the lowest resolved energy transfer and the upper limit
$\hbar \omega_2 \lesssim$1meV restricts the low-$\omega$ region of interest. In (\ref{a1}) the
dynamical structure factor has a familiar form
%02
\begin{equation}
S\left({\bf q}, \omega; T\right) = \left(1-e^{-\hbar \omega
/k_{B}T}\right)^{-1}\mbox{Im}\chi\left({\bf q}, \omega; T\right), \label{a2}
\end{equation}
where $\chi\left({\bf q}, \omega; T\right)$ is the dynamic spin susceptibility. For this we
choose a single-pole description
%03
\begin{equation}
\chi\left({\bf q}, \omega; T\right)\simeq \frac{\chi\left({\bf q};
T\right)}{1-i\omega/\Gamma_{\bf q}(T)}; \label{a3}
\end{equation}
where $\chi\left({\bf q}; T\right)$ and $\Gamma_{\bf q}\left(T\right)$ are the temperature
dependent static spin susceptibility and the spin relaxation rate, respectively. It is worth emphasizing here that a single Lorentzian spectral form of nearly critical spin fluctuations  corresponding to the Eq.(\ref{a3}) provides an adequate description of low-temperature INS data~~\cite{Krimmel99,Lee01,Murani04} in LiV$_2$O$_4$.

Our previous study~\cite{Yushankhai07} of spin fluctuations in strongly correlated itinerant
electron system LiV$_2$O$_4$ was carried out at $T= 0$ within the RPA approach based on the
realistic electronic band structure of this material. 
The theory suggests that  the  spinel LiV$_2$O$_4$ is near to a magnetic instability 
and possesses  a rather unusual  paramagnetic ground state:  The pronounced
low-$\omega$  spin fluctuations are located in ${\bf q}$-space in the vicinity of the "critical"
${\bf Q}_c$-surface  with a mean radius  $Q_c  \simeq$ 0.6 $\AA^{-1}$,
Fig.7 in Ref.[\onlinecite{Yushankhai07}]. The suggested strong degeneracy
of "critical" wave vectors  ${\bf Q}_c$ means that on approaching a magnetic instability the system cannot
choose a unique wave vector of a magnetic structure which minimize the free energy of spin fluctuations.
Instead, it is frustrated between different structures with different wave vectors and equally low free energy.

For powder samples of LiV$_2$O$_4$ measured~\cite{Lee01,Murani04} at
$T\lesssim 2$K, all the "critical" spin fluctuations  at $|{\bf q}| \sim Q_c$ contribute  because of 
the angle averaging at a given $|{\bf q}|$ and the
corresponding low-$\omega$ scattering intensity $I\left(Q_c; T\to 0\right)$   dominates;  for
instance $I\left(Q_c; T\to 0\right)/I\left( q; T\to 0\right)\gg 1$, where small values of $q$
around the BZ center are implied. As mentioned above, experiment shows that with
increasing temperature  the small $q$ intensity $I\left( q; T\right)$ grows fast and for
$T\gtrsim 60$K tends to be dominant. So far, a detailed explanation  for the reversal of  the
ratio $I\left(Q_c; T\right)/I\left( q; T\right)$ with increasing $T$ is still lacking.

The purpose of the present work is two-fold. First, we  aim to give an explanation for the observed
shift of the INS intensity under warming. Our argumentation is
based on the markedly different temperature evolutions of  "critical" spin fluctuations  at   $|{\bf q}|\sim Q_c$
and those at small ${\bf q}$. Second, 
to extend our previous RPA study of  spin fluctuations at $T=0$  in  LiV$_2$O$_4$ to finite temperatures,
in the present work one step beyond the RPA theory is made. In the extended theory, known as the   
self-consistent renormalization (SCR) theory of spin fluctuations,~\cite{Moriya85,Moriya73,Hasegawa74}
leading corrections to  the inverse RPA spin susceptibility $\chi^{-1}_{RPA}\left({\bf Q}_c\right)$ are included to involve effects of spin fluctuation interactions in a self-consistent manner. One aspect that distinguishes considerably the SCR  theory
applied   here for   LiV$_2$O$_4$   from 
the earlier applications of this theory to other electronic systems near magnetic instabilities has to be  especially emphasized. 
As suggested in~\cite{Yushankhai07} and outlined above, it is a rather unusual distribution in ${\bf q}$-space  of slow "critical" spin fluctuations that dominate  the  paramagnetic  behavior of  LiV$_2$O$_4$ in the limit $T\to 0$.

Based on  the present SCR theory,  a  phenomenological description for the dynamical spin susceptibility   
$\chi\left({\bf Q}_c,\omega;T\right)$  measured by INS~\cite{Krimmel99,Lee01,Murani04}  is developed.
  As shown below,  the properly parametrized SCR theory is capable in giving at low temperatures a satisfactory fit to the experimentally observed $T$-dependences of both the static  susceptibility 
 $\chi\left({\bf Q}_c; T\right)$ and the relaxation rate  $\Gamma_Q\left(T\right)$ for the "critical"   spin fluctuations.

%%%%%%%%%%%%%%%%%%%%%%%% --- II ---%%%%%%%%%%%%%%%%%%%%%%%%%

\section{Basic equations of the SCR theory }

Focusing on the wave vector region at   ${\bf q}\sim {\bf Q}_c$, where spin fluctuations are highly enhanced at low $T$,
within the SCR theory
a temperature dependence of the static spin susceptibility  $\chi\left({\bf q}; T\right)$ is
described by the following equation~\cite{Moriya85} ($\hbar =1, k_{B} = 1$ ):

\begin{eqnarray}
&&\frac{1}{\chi\left({\bf q}; T\right)}=\frac{1}{\chi_0\left({\bf
q}\right)}-2U\nonumber\\
&&\hspace{2mm}+\frac{5}{3}\mathcal{F}_{\bf
Q}\frac{1}{N}\int_{-\infty}^{+\infty}\frac{d\omega}{2\pi} \coth\left(\frac{
\omega}{2T}\right)\sum_{\bf{q}^\prime}\mbox{Im} \chi\left({\bf q}^\prime, \omega; T\right)
\label{a4}
\end{eqnarray}
Here $\chi_0\left({\bf q}\right)$ is the static susceptibility for non-interacting electrons.
The term $-2U$
takes into account electron  correlations in the RPA approximation, where the parameter  $U>0$ is the on-site electron repulsion in an effective Hubbard model.  Thus, first two terms on the
right-hand side  of Eq.(\ref{a4}) give the inverse RPA spin susceptibility $\chi^{-1}\left({\bf
q}\right)$.  A weak  $T$-dependence of
 $\chi^{-1}\left({\bf q}\right)$ is brought about by the Fermi  distribution function entering the generalized
Lindhard function $\chi_0\left({\bf q}\right)$.  In the low-$T$ range of interest,  the
corresponding temperature corrections are controlled by  the extremely small quantity
$\left(T/\epsilon_F\right)^2$,  where $\epsilon_F$ is the Fermi energy, and thus    can be neglected. Primarily,
a temperature variation of $\chi\left({\bf q}; T\right)$ is induced by the last term on the
right-hand side  of Eq.(\ref{a4}) due to mode-mode coupling of spin fluctuations. The coupling strength  is given by the  
constant  $\mathcal{F}_{\bf Q}$.  The 
spectral intensity $\mbox{Im} \chi\left({\bf q}, \omega; T\right)$  is   dominated by spin fluctuations  at   ${\bf q}\sim {\bf Q}_c$  characterized by $\chi\left({\bf Q}_c; T\right)$ and  
a set of complementary 
parameters.  The procedure is developed  in the next section, where Eq.(\ref{a4}) at  ${\bf q}={\bf Q}_c$ takes a form of a parametrized integral equation for  $\chi\left({\bf Q}_c; T\right)$.

With the use of the decomposition
$\coth(\omega/2T)=1+2f_B(\omega/T)$, where $f_B(\omega/T)$ is the Bose distribution function,
the last term in Eq.(\ref{a4})  can be split  into two parts. The first gives a
contribution from zero point fluctuations with the main effect of renormalizing  the parameter
$U\to U_{eff}$. The second part involving $f_B(\omega/T)$ gives  the explicit and dominant
$T$-dependence of  $\chi\left({\bf q}; T\right)$; at $T=0$ this contribution to $\chi\left({\bf
q}; T\right)$ is zero. Therefore,
\begin{equation}
\frac{1}{\chi\left({\bf q}; T=0\right)}= \frac{1-2U_{eff}\chi_0({\bf q})}{\chi_0\left({\bf
q}\right)} \label{a5}
\end{equation}
Now  Eq.(\ref{a4}) can be rewritten as
\begin{eqnarray}
&&\frac{1}{\chi \left({\bf q}; T\right)}=\frac{1}{\chi \left({\bf q}; T=0\right)} \nonumber\\
&&\hspace{2mm}+{\bar\mathcal{F}}_{\bf Q}\int_{0}^{+\infty}\frac{d\omega}{2\pi} \frac{1}{e^{ \omega /
T} -1} \frac{1}{N} \sum_{\bf{q}^\prime}\mbox{Im} \chi\left({\bf q}^\prime, \omega; T\right),
\label{a6}
\end{eqnarray}
where ${\bar\mathcal{F}}_{\bf Q}=\left(20/3\right) \mathcal{F}_{\bf Q}$; $N$ is the number of
primitive cells in the sample volume. The ${\bf q}$-summation is over the BZ of the fcc lattice
inherent to the pyrochlore lattice of the magnetic V-ions in the spinel structure of  LiV$_2$O$_4$. Hereafter, $\chi\left({\bf q}; T\right)$ means
the spin susceptibility calculated per primitive cell (4 V-atoms).

In Eq.(\ref{a6}) the integral quantity
\begin{equation}
\lambda\left(T\right) = \int_{0}^{+\infty}\frac{d\omega}{2\pi} \frac{1}{e^{ \omega /
T} -1} \frac{1}{N} \sum_{\bf{q}^\prime}\mbox{Im} \chi\left({\bf q}^\prime, \omega; T\right),
\label{a7}
\end{equation}
has,  up to a constant factor, a meaning of the mean square amplitude of the thermally induced spin fluctuations;   $\lambda\left(T\right)$ is a monotonically growing function of $T$ with the property $\lambda\left(T=0\right)=0$.

At the next step,  the imaginary part of
 the dynamic spin susceptibility $\chi\left({\bf q}, \omega; T\right)$ entering  Eq.(\ref{a6})
and  a distribution in ${\bf q}$-space of dominant spin fluctuations  have to 
be specified with more detail.

%%%%%%%%%%%%%%%%%%%%%%%%--- III ---- %%%%%%%%%%%%%%%%%%%%%%

\section{"Critical" spin fluctuations  in  LiV$_2$O$_4$}

 In  INS measurements~\cite{Lee01, Murani04} on powder samples of   LiV$_2$O$_4$ ,  highly enhanced spin fluctuations
 were detected at    "critical" wave vectors of the length $|{\bf Q}_c|\sim$0.6 $\AA^{-1}$.   However,  a direction of  
 ${\bf Q}_c$ remained unknown. 
 In a subsequent
 theoretical study~\cite{Yushankhai07}, the dynamical spin susceptibility $ \chi\left({\bf q}, \omega; T\right)$ was calculated
at $T=0$  for the realistic band structure of the itinerant electron paramagnet  LiV$_2$O$_4$ with local on-site electron interactions
treated in the RPA approach. The calculations  of  $ \chi\left({\bf q}, \omega; T\right)$  performed along
high-symmetry directions in ${\bf q}$-space revealed that for each of the tested   ${\bf q}$-directions,  strongly enhanced spin fluctuations
occur at  a  wave vector ${\bf Q}_i$ of  the length close to the experimentally measured "critical" one, 
$|{\bf Q}_i|\sim |{\bf Q}_c|\sim$0.6 $\AA^{-1}$.
The end
points of the calculated wave vector manifold  $\{{\bf Q}_i\}$ can be viewed as lying on a closed surface
called~\cite{Yushankhai07} the  "critical"  ${\bf Q}_c$-surface. It can be approximated with a
polyhedron surface formed by edge-sharing polygons in such a way that the end point of a
particular ${\bf Q}_i$ is the $i$-th polygon center.
Below the manifold of ${\bf Q}_i$ is denoted as ${\bf Q}_c=\{{\bf
Q}_i\}$ and a prescription ${\bf q}={\bf Q}_c$ would mean that any of ${\bf Q}_i$ can be taken
for ${\bf q}$. 

The sum over BZ entering the expressions (\ref{a6}) and (\ref{a7}) can be now expanded in the
following way
\begin{equation}
\sum_{\bf q}\mbox{Im}\chi\left({\bf q}, \omega; T\right)=\sum_{i} \sum_{{\bf q}_i}\mbox{Im}
\chi\left({\bf Q}_i+{\bf q}_i, \omega; T\right),  \label{a9}
\end{equation}
where each ${\bf q}_i$-summation is over the ${\bf q}$-states inside the $i$-th 3D wedge
associated with a wavevector ${\bf Q}_i$. The origin   ${\bf q}_i=0$ is placed at the center of
the  $i$-th polygon on the "critical"  surface.  The ${\bf q}_i$-regions near the ${\bf
Q}_c$-surface are characterized by "critical",  i.e. strongly enhanced and slow  spin   fluctuations
which provide the main contribution to $\lambda\left(T\right)$, Eq.(\ref{a7}).

With the set of normals   $\{{\bf n}_i\} = \{{\bf Q}_i/ |{\bf Q}_i|\}$ to the ${\bf
Q}_c$-surface, one may write down ${\bf q}_i=q_i^{||}{\bf n}_i + {\bf q}_i^{\bot}$; near ${\bf
Q}_c$ the  two-component vectors ${\bf q}_i^{\bot}$ are confined to the $i$-th
polygon. We suggest the following expansion of the dynamic spin susceptibility for low $\omega$
and near the  ${\bf Q}_c$-surface:
\begin{eqnarray}
&&\frac{1}{\chi\left({\bf Q}_i+{\bf q}_i, \omega; T\right)}\nonumber\\
&=&\frac{1}{\chi\left({\bf Q}_c; T\right)}
+ A\left(q_i^{||}\right)^2 + B\left({\bf q}_i^{\bot}\right)^2 -iC\omega, \label{a10}
\end{eqnarray}
where $B\ll A$. The parameters $A,B$ and $C$ are assumed  to be $T$-independent in the low-$T$
region where the SCR theory is valid. The expansion (\ref{a10}) is compatible with our previous
study~\cite{Yushankhai07}  and presents a further development of the model along the way proposed
there. A peculiar property of the suggested model of "critical" spin
fluctuations in  LiV$_2$O$_4$ is  their strongly anisotropic character: the dispersion ($\sim 1/A$)
in the direction parallel to the normals $\{{\bf n}_i\}$ to the ${\bf Q}_c$-surface is much
smaller than that ($\sim 1/B$) in the perpendicular directions. This property is verified in the next section
by showing that a better fit of the calculated model results to INS experimental data is achieved 
with the anisotropy parameter $b=B/A$ tending to zero.

Let us consider Eq.(\ref{a6}) taken for ${\bf q}={\bf Q}_c$.  Note that  the
coupling constant  ${\mathcal{F}}_{\bf Q}$ is assumed to be degenerate in the set of $\{{\bf Q}_i\}$
and, hence, can be denoted as  ${\mathcal{F}}_Q$.
Therefore, any  $\chi\left({\bf Q}_i; T\right)$  from the  ${\bf Q}_c=\{{\bf Q}_i\}$ manifold
obeys the same equation. With the use of (\ref{a9}) and (\ref{a10}), this leads to the explicit
equation for $\chi\left(Q_c; T\right)$ that must be solved self-consistently. 

 In the
present form, the SCR theory is parametrized with five parameters, $ \chi\left(Q_c;T=0\right),  A, B, C$
 and  ${\mathcal{F}}_Q$.  At the final stage, we put the theory on a quantitative ground by 
adjusting the parameter values when comparing  the calculated model results with 
experimental  INS data~\cite{Lee01} for the spin susceptibility in  LiV$_2$O$_4$.

To be close to the standard notation of the SCR theory~\cite{Moriya85}, we introduce, instead of
$A$ and $C$,  the following parameters
\begin{equation}
T_A=\frac{Aq_B^2}{2},\hspace{5mm}T_0=\frac{Aq_B^2}{2\pi C}, \label{a11}
\end{equation}
where $q_B$ is the effective radius of the BZ boundary given in terms of a primitive cell volume
$v_0$ as $q_B=\left(6\pi^2/v_0\right)^{1/3}$. Next,  the reduced inverse susceptibility at ${\bf
q}={\bf Q}_c$ is defined as
\begin{equation}
y_Q\left(T\right)=\frac{1}{2T_A\chi\left(Q_c; T\right)}. \label{a12}
\end{equation}
With these notation one obtains
\begin{eqnarray}
&&\mbox{Im}\chi({\bf Q}_i+{\bf q}_i, \omega; T)
=\frac{1}{2T_A}(\omega/2\pi T_0)\nonumber\\
&&\times\left(
\left\{y_Q(T) + (\frac{q_i^{||}}{q_B})^2  
+ b\hspace{1mm}(\frac{{\bf q}_i^{\bot}}{q_B})^2\right\}^2 
+ (\omega/2\pi T_0)^2\right)^{-1}
%\nonumber\\
%&&\times\frac{\omega/2\pi
%T_0}{\left[ y_Q\left(T\right) + \left(q_i^{||}/q_B\right)^2  + b\left({\bf
%q}_i^{\bot}/q_B\right)^2\right]^2 + \left(\omega/2\pi T_0\right)^2},
\label{a13}
\end{eqnarray}
where $b=B/A$.

Below, when performing in (\ref{a9}) the summation over ${\bf q}_i$ and $i$, two dimensionless
cutoff numbers, $z_c=(q_c^{||}/q_B)$ and $x_c=(q_c^{\bot}/q_B)^2$, are introduced. For the former
we take $z_c\approx 1/2$,  which distinguishes the region of "critical" spin fluctuations from that
with small $q$ ones. The latter  has a meaning of a square  dimensionless {\it mean} radius of the polygons
forming the ${\bf Q}_c$-surface. This can be related to the ${\bf Q}_c$-surface area: $S_Q=\sum_i
S_{Q,i}\approx \pi q_B^2\sum_i x_c$. In the spherical approximation, $S_Q=4\pi Q_c^2$, one has
$\sum_i x_c \approx 4\left(Q_c/q_B\right)^2$.

By inserting the expressions (\ref{a13}) into (\ref{a9}) and (\ref{a7}), we get
\begin{eqnarray}
&&\lambda\left(T\right) = \frac{3T_0}{4T_A}\sum_{i}
\int_{0}^{z_c}dz_i\int_{0}^{x_c}dx_i\int_{0}^{\infty} d\nu \nonumber\\
&&\hspace{5mm}\times\frac{\nu}{e^{2\pi\nu}-1}
\frac{t^2}{\left[ y_Q\left(t\right) + z_i^2 + bx_i\right]^2 + \left(\nu
t\right)^2},
 \label{a14}
\end{eqnarray}
where $t=T/T_0$. First, by performing the integration over $x_i$, one obtains
\begin{eqnarray}
&&\int_{0}^{x_c}dx_i\frac{1}{\left[ y_Q\left(t\right) + z_i^2 + bx_i\right]^2 +
\left(\nu t\right)^2} \\
&&=\frac{1}{b\nu t}\left(\tan^{-1}\frac{\nu t}{y_Q\left(t\right) +
z_i^2} - \tan^{-1}\frac{\nu t}{y_Q\left(t\right) + z_i^2 + bx_c}\right).
\nonumber
 \label{a14}
\end{eqnarray}
At the next step, the $\nu$-integration in (\ref{a14}) is performed straightforwardly by
recalling that
\begin{eqnarray}
&&\int_{0}^{\infty}d\nu \frac{\tan^{-1}\left(\nu/y\right)}{e^{2\pi\nu}-1}
\nonumber\\
&&=\frac{1}{2}\left\{\ln \Gamma\left(y\right)-
\left(y-\frac{1}{2}\right)\ln y + y - \frac{1}{2}\ln 2\pi\right\},
 \label{a15}
\end{eqnarray}
where $\Gamma\left(y\right)$ is the gamma function. In (\ref{a14}), a value of the remaining
integral over $z_i$, where the subscript $i$ denotes a polygon number, does not depend on $i$. In
the resulting integral expression for $\lambda$ the $i$-summation enters as a common factor
$\sum_i x_c \approx 4\left(Q_c/q_B\right)^2$.

Finally, we arrive at the following equation for the reduced inverse susceptibility~\cite{Kondo02}
\begin{eqnarray}
&&y_Q\left(t\right)=y_Q\left(0\right)
+g_Q\int_{0}^{z_c}dz\nonumber\\
&&\times\frac{\phi\left(\left\{y_Q(t) +z^2\right\}/t\right)
-\phi\left(\left\{y_Q(t) + z^2 +bx_c\right\}/t\right)}{bx_c /t},
 \label{a16}
\end{eqnarray}
with
\begin{eqnarray}
\phi\left(u\right) &=&\ln  \Gamma\left(u\right)-\left(u-\frac{1}{2}\right)\ln u +u -\frac{1}{2}\ln 2\pi, \nonumber\\
g_Q &=& \frac{5T_0}{T_A^2}\left(\frac{Q_c}{q_B}\right)^2 {\mathcal F}_Q, \hspace{5mm} z_c\simeq \frac{1}{2}.
 \label{a17}
\end{eqnarray}
The choice for $z_c$ in (\ref{a17}) is justified earlier.  Provided the parameter values of   $y_Q\left(0\right)$,  $g_Q$  and  $bx_c$
are fixed as discussed below,  the appearance of a solution of Eq.(\ref{a16}) for $y_Q\left(t\right)$ as a function of $t$  is checked to be qualitatively insensitive to a variation of   $z_c$.

Beside the basic equation (\ref{a16}),  the present  SCR theory includes a set of five parameters which
 are now denoted as  $y_Q\left(0\right)$,  $T_A$,  $T_0$,  $g_Q$  and  $bx_c$.  The parameters  $T_A$ and  $T_0$ characterize,
 at $T\to 0$, the momentum and frequency  spread of "critical" spin fluctuations,  $g_Q$  is the effective mode-mode coupling constant
 and   $bx_c$ is a measure of the anisotropy of the spin fluctuation dispersion in ${\bf q}$-space.
Given that the SCR theory provides an expected reasonable approach to  a description of  INS data~\cite{Krimmel99,Lee01,Murani04},
the parameter values can be safely estimated with a fit procedure.

%%%%%%%%%%%%%%%%%%%%%%%%%%%%%%%%%%%%%%%%%%%%%%%%%%%%%%%%

\section{Temperature renormalization of "critical"  spin fluctuations}

In a single-pole approximation (\ref{a3})  to the dynamic spin susceptibility,  temperature evolution of "critical" spin fluctuations
is entirely described by $T$-dependences of the static spin susceptibility $\chi\left(Q_c; T\right)$ and the spin relaxation rate
 $ \Gamma_{ Q} \left(T\right)$.  Our aim now is to give a phenomenological description of  $\chi\left(Q_c; T\right)$ and 
 $ \Gamma_{ Q} \left(T\right)$  measured in INS experiment~\cite{Lee01,Murani04} in terms of the parametrized SCR theory
developed in the proceeding sections.  

Let us first note that the expansion (\ref{a10}) 
can be rewritten in the form (\ref{a3}), which leads to the following relations
\begin{eqnarray}
&&\chi\left({\bf Q}_c+{\bf q}; T\right)\nonumber\\
&=&\frac{\chi\left(Q_c; T\right)}{1+\left[q^{||}/\kappa_Q\left(T\right) \right]^2 + 
b\left[ {\bf q}^{\bot}/\kappa_Q\left(T\right) \right]^2},
\label{a20}
\end{eqnarray}
\begin{eqnarray}
&&\Gamma_{{\bf Q}+{\bf q}} \left(T\right)\nonumber\\
&=& \Gamma_{ Q} \left(T\right)\left(1+\left[q^{||}/\kappa_Q\left(T\right) \right]^2 + 
b\left[ {\bf q}^{\bot}/\kappa_Q\left(T\right) \right]^2\right),
\label{a21}
\end{eqnarray}
where
\begin{equation}
\kappa_Q^2\left(T\right)=\frac{1}{A\chi\left(Q_c; T\right)}=q_B^2y_Q\left(T\right),
\label{a22}
\end{equation}

\begin{equation}
\Gamma_{ Q} \left(T\right)=\frac{A}{C}\kappa_Q^2\left(T\right)=2\pi T_0y_Q\left(T\right).
\label{a23}
\end{equation} 
By using  the  INS estimates~\cite{Lee01} for the inverse  correlation length $\kappa_Q\left( T\to 0\right)=\xi^{-1}\sim$0.16 $\AA^{-1}$ 
 (at $Q_c\simeq $0.6 $\AA^{-1}$)  and the spin relaxation rate $\Gamma_{ Q} \left(T\to 0\right)\simeq 1.4$meV,
and recalling that $q_B=$0.76 $\AA^{-1}$, one obtains from Eqs.(\ref{a21}) and (\ref{a22}) the following parameter values
\begin{equation}
y_Q\left(0\right)\simeq 0.044, \hspace{5mm} T_0\simeq 60 K.
\label{a24}
\end{equation} 

Next, having the ratio  $\chi\left(Q_c; 0\right)/\chi\left({\bf q}\to 0; 0\right)\simeq$4 reported in Ref.[\onlinecite{Lee01}] and the estimate
$\chi\left({\bf q}= 0; T\to 0\right)\simeq$0.15/meV ( per primitive cell  and  $2\mu_B=$1 implied) derived from Ref.[\onlinecite{Kondo99}]  one obtains
\begin{equation}
T_A=\frac{1}{2y_Q\left(0\right)\chi\left(Q_c; 0\right)}\simeq 220 K.
\label{a25}
\end{equation}
Two remaining parameters,  $g_Q$ and $bx_c$, can be estimated from a fit of experimentally
observed temperature variations of  $\chi\left(Q_c; T\right)/\chi\left(Q_c; 0\right)$ and  
$\Gamma_{ Q} \left(T\right)/\Gamma_{ Q} \left(0\right)$ to a solution $y_Q\left(t\right)$ of Eq.(\ref{a16}).

%%%%%%%%%%%%%%%%%%%%%%%%%% FIGURE %%%%%%%%%%%%%%%%%%%%%%%%%%%%%%%%%%%%%
\begin{figure}[th]
\begin{center}
\resizebox{75mm}{!}
{\includegraphics{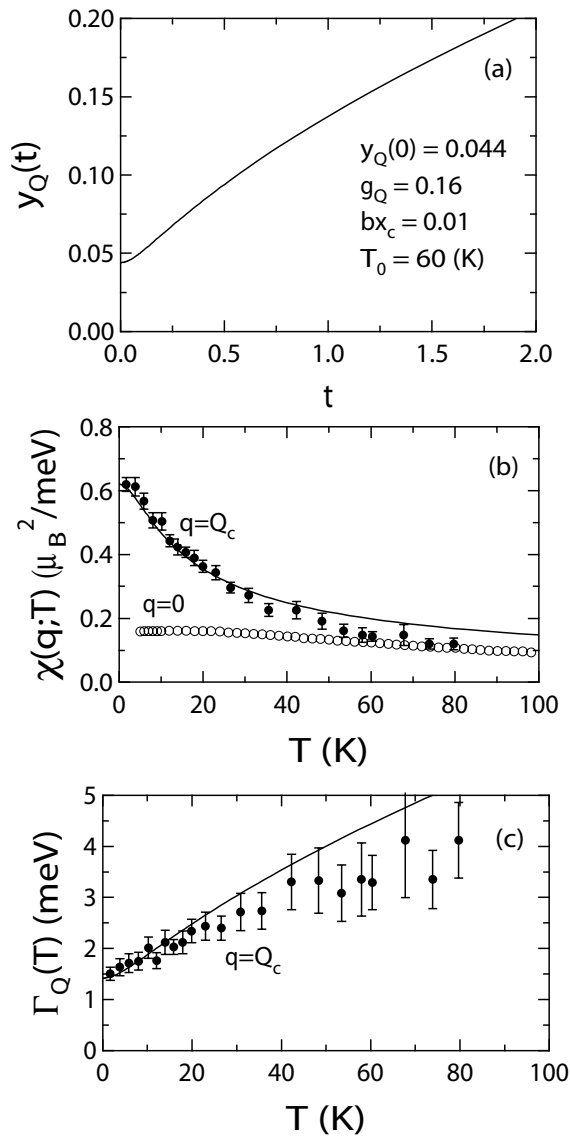}}
\end{center}
\caption{
(a) The solution of Eq. (\ref{a16})  for the reduced inverse static spin susceptibility at $q = Q_c$
as a function of the reduced temperature $t=T/T_0$;  the full set of fit parameters are  given in section IV;
(b) Static spin susceptibilities at $q = Q_c$ (solid circles) and $q = 0$ (open circles) as functions of temperature
 observed in INS and magnetic measurements on  LiV$_2$O$_4$.  The solid line is a fit to $\chi\left(Q_c; T\right)$ using the same solution as in (a);
(c) Spin relaxation rate at $q = Q_c$  as a function of temperature observed in INS experiment. The solid line is 
a fit using again the  solution of (a).
}
\end{figure}
%%%%%%%%%%%%%%%%%%%%%%%%%%%%%%%%%%%%%%%%%%%%%%%%%%%%%%%%%%%%%%%%%%%%%%%

In  Fig.1, the solution for  $y_Q\left(t\right)$  with $g_Q=0.16$ and  $bx_c=0.01$,
and the corresponding fit  to  INS experimental data  for 
 $\chi\left(Q_c; T\right)$ and  $\Gamma_{Q} \left(T\right)$ are plotted.  Provided  the values of  $y_Q\left(0\right)$ and
 $T_0$   are kept as in  (\ref{a24}),  variation of   $g_Q$ and  $bx_c$ around the selected values, 
$g_Q=0.16$ and  $bx_c=0.01$,  would lead to a smooth modification of a curvature of the solution  $y_Q\left(t\right)$
presented in the upper panel of Fig.1. With respect to the anisotropy parameter $bx_c$,  we found a better coincidence,
mostly for  $T\gtrsim 30$K, between the theory and experiment  by sending $b$ to smaller values  and we adopt the value  $bx_c=0.01$ as the most representative one.

Thus,  with  $g_Q$ and  $bx_c$ values indicated above,  the best  overall fit  to experimental data for $\chi\left(Q_c; T\right)$ and  $\Gamma_{ Q} \left(T\right)$ is achieved.
As expected, a better agreement between the theory   and experiment  is reached  in the low temperature region, 
$T\lesssim 30$K, while   for  $T\gtrsim 30$K only main trends in $T$-dependences of  $\chi\left(Q_c;T\right)$ and $\Gamma_Q\left(T\right)$ are reproduced  by the present  theory. Actually,   
  "critical" spin fluctuations at  $|{\bf q}|\sim Q_c$, being dominant  at  $T\to 0$,  get suppressed  with increasing $T$.
This means that, as $T$  becomes high enough, 
the  spectral  density $\sim\mbox{Im}\chi\left({\bf q}, \omega; T\right)$  integrated over the whole BZ,  Eq.(\ref{a6}), is no more dominated
 by  fluctuation modes at $q\sim Q_c$ only, and  contributions of other modes,  more likely those at small $q$,  have to be involved
in a more complete theory.
%%%%%%%%%%%%%%%%%%%%%%%%%%%%%%%%%%%%%%%%%%%%%% 
\section{Temperature redistribution of spin fluctuation spectral intensity}

To compare  the INS  intensities at  ${\bf q}\sim {\bf Q}_c$  and   ${\bf q}\to 0$ and their evolution with temperature,
let us start with examining  of  $T$-dependences  of  $\chi\left({\bf q};T \right)$ and $\Gamma_{\bf q}\left( T\right)$ entering
the expression (\ref{a3})  for the small-${\bf q}$ and  low-$\omega$ dynamic spin susceptibility.
First, one may relate the small-${\bf q}$ quantities $\chi\left({\bf q};T \right)$ and $\Gamma_{\bf q}\left( T\right)$  in
a similar way as done in preceding section for ${\bf q}\sim {\bf Q}_c$, with the use of  the following expansion
\begin{equation}
\frac{1}{\chi\left({\bf q}, \omega; T\right)}=\frac{1}{\chi\left(0; T\right)}
+ A_0 q^2  -i\omega C_0/q,
\label{a26}
\end{equation}
The extra factor $1/q$ in the last term of Eq.(\ref{a26}) arises since the uniform magnetization is a constant of motion.
Then one obtains
\begin{equation}
\chi\left({\bf q}; T\right)=\frac{\chi\left(0; T\right)}{1+\left[q/\kappa_0\left(T\right) \right]^2 },
\label{a27}
\end{equation}
\begin{equation}
\Gamma_{{\bf q}} \left(T\right) \simeq \frac{A_0}{C_0}\kappa_0^2\left(T\right) q,
\label{a28}
\end{equation}
where
\begin{equation}
\kappa_0^2\left(T\right)=\frac{1}{A_0\chi\left(0; T\right)}.
\label{a29}
\end{equation}
The INS~\cite{Lee01} and magnetic measurements~\cite{Kondo99} show that  low-$T$ variations of spin susceptibilities  $\chi\left(0; T\right)$  and 
$\chi\left(Q_c; T\right)$  are characterized by dramatically different scales. Actually,
as temperature increases from $T\sim 1$K up to $T\sim 60$K,  the value of $\chi\left(Q_c; T\right)$ drops
by a factor of 4, while  about a twenty percent decrease in  $\chi\left(0; T\right)$ is observed only as seen in Fig.1.

These preliminaries enable us now  to give an account for a temperature variation of the
ratio  $I\left(Q_c; T\right)/I\left( q; T\right)$.  Here, the integrated low-$\omega$
INS intensities are defined as in  Eqs.(\ref{a1}) and (\ref{a2}),  and  small, but  finite
wave vectors ${\bf q}$  near the  BZ center  are implied in $I\left( q; T\right)$.

Both for $q\sim Q_c$ and small $q$,  in the low-$\omega$ limit the imaginary part of the dynamic
spin susceptibility (\ref{a3}) can be approximated as
$\mbox{Im}\chi\left({\bf q}, \omega; T\right) \approx \chi\left({\bf q}; T\right)\omega/\Gamma_{\bf q}\left(T\right)$,
and for finite temperatures, $T>\omega$,   the temperature-balance factor in (\ref{a2}) can be approximately
replaced as follows:  $\left(1-e^{-\omega/T}\right)^{-1}\sim T/\omega$.
Then, with the use of  (\ref{a20})-(\ref{a23})  and  (\ref{a27})-(\ref{a29}) one obtains
\begin{equation}
\frac{I\left(Q_c; T\right)}{I\left( q; T\right)}\approx \frac{\Gamma_q\left(T\right)}{\Gamma_Q\left(T\right)}
\frac{\chi\left(Q_c; T\right)}{\chi\left(q; T\right)}\approx \frac{qC}{C_0}\left[\frac{\chi\left(Q_c; T\right)}{\chi\left(0; T\right)}\right]^2.
\label{a30}
\end{equation}
For $T\ll T_0=60$K, the insertion of the experimentally determined  ratio $\chi\left(Q_c; T\right)/\chi\left(0; T\right)\sim \chi\left(Q_c; T\to 0\right)/\chi\left(0; T\to 0\right) \approx 4$ 
into Eq.(\ref{a30})  yields   $I\left(Q_c; T\right)/I\left( q; T\right)\gg 1$,  provided  $qC/C_0\sim 1$. Further, as $T$ is
elevated gradually,  the decrease of $\chi\left(Q_c; T \right)$, together with a relatively weak variation of $\chi\left(0; T\right)$,
is resulted for the last term in Eq.(\ref{a30}) in a fast decreasing function of $T$ such that 
$\left[\chi\left(Q_c; T\right)/\chi\left(0; T\right)\right]^2\approx 1$ at $T\sim 60$K.  Thus, the ratio 
$I\left(Q_c; T\right)/I\left( q; T\right)$ is reduced by almost a factor of 16.

The results of the this section can be comprehended  as follows.
For small ${\bf q}$, the low-$\omega$ scattering
intensity increases with $T$  mainly due to the thermal-balance factor entering the expression
(\ref{a2})  because the static spin susceptibility  $\chi\left({\bf q}; T\right)$ and the
relaxation rate $\Gamma_{\bf q}\left(T\right)$  as functions of $T$ are  slowly varying observables  for
$T\lesssim 60$K. In contrast,  at ${\bf q}\simeq {\bf Q}_c$ the
thermal-balance factor increase is cancelled out because of ({\it i} ) a fast  decrease with $T$ of the static
spin suceptibility  $\chi\left({\bf Q}_c; T\right)$  and  ({\it ii} ) a concomitant shifting
of the spectral intensity  to higher frequencies. The latter means a fast
increase with $T$ of the spin relaxation rate $\Gamma_{\bf Q}\left(T\right)$ measured at  ${\bf
q}\simeq {\bf Q}_c$, which is in our analysis related to a fast temperature decrease of
$\chi\left({\bf Q}_c; T\right)$. This markedly distinct temperature behavior of spin fluctuations located
in different regions of ${\bf q}$ space manifests itself in a shift of the dominant low-$\omega$  INS intensity 
from $q\sim Q_c$ at $T\sim 1$K  to  small $q$ near the BZ center for $T\gtrsim 60$K.

%%%%%%%%%%%%%%%%%%%%%%%%%%%%%%%%%%%%%%%%%%%%%%%%%%%
\section{Conclusion}
Like in many strongly correlated metallic systems, including heavy fermion compounds and high-$T_c$
cuprates, that are close to magnetic instabilities at  $T\to 0$,  the pronounced slow spin fluctuations are suggested
to influence considerably low-$T$ properties of the paramagnetic spinel  LiV$_2$O$_4$ as well.
In the present work one step beyond the RPA theory was made  and effects of spin fluctuation interactions were involved in the  form of the SCR theory.  
On this ground,  a phenomenological description of the experimentally observed  temperature variation of the  dynamic spin susceptibility
$\chi\left(Q_c, \omega; T\right)$ for the dominant "critical" spin fluctuations  was developed. 

A special feature of LiV$_2$O$_4$  which distinguishes   the present
SCR theory  from its earlier applications to other electronic systems near magnetic instabilities is a peculiar distribution
 in ${\bf q}$ space of low-$\omega$ "critical" spin fluctuations  dominating the paramagnetic state of LiV$_2$O$_4$ in the limit $T\to 0$.  
 As shown,  the properly parametrized SCR theory is in agreement with  the experimentally observed  low-$T$ variations of the static spin susceptibility  $\chi\left(Q_c; T\right)$
and the relaxation rate $\Gamma_Q\left(T\right)$ for the "critical" spin fluctuations.  

Based on the markedly different temperature evolutions of  "critical" spin fluctuations and those at small ${\bf q}$,
we gave an explanation for the warming shift of the INS intensity from the  "critical"  region  $|{\bf q}|\sim Q_c$ to
the one of smaller ${\bf q}$.

On the future perspective,  we remark the following.
In order to check the full  consistence of the  spin fluctuation theory developed in the present paper,  one needs to examine
whether the measured temperature dependences of the heat capacity $C(T)$,  the electrical resistivity $\rho (T)$
and the NMR spin-lattice relaxation rate $T_1^{-1}(T)$ are also in accordance with the theory.

%in~\cite{Krimmel99,Lee01,Murani04}. 
%%%%%%%%%%%%%%%%%%%%%%%%%%%%%%%%%%%%%%%%%%%%%%%%%
{\bf Acknowledgements}\\

V.Yu. acknowledges a  support  from  the project  SFB 463 by Deutsche Forschungsgemeinschaft.
%%%%%%%%%%%%%%%%%%%%%%%%%%%%%%%%%%%%%%%%%%%%

%\newpage

%\begin{center}
%\Large{References}
%\end{center}

%%%%%%%%%%%%%%%%%%%%%%%%%%%%%%%%%%%%%%

%\begin{center}
%\Large{Figure caption}
%\end{center}
%Fig.1 \\
%(a) The solution of Eq. (\ref{a16})  for the reduced inverse static spin suscep%tibility at $q = Q_c$
%as a function of the reduced temperature $t=T/T_0$;  the full set of fit parame%ters are  given in section IV;\\
%(b) Static spin susceptibilities at $q = Q_c$ (solid circles) and $q = 0$ (open% circles) as functions of temperature
% observed in INS and magnetic measurements on  LiV$_2$O$_4$.  The solid line is% a fit to $\chi\left(Q_c; T\right)$ using the same solution as in (a);\\
%(c) Spin relaxation rate at $q = Q_c$  as a function of temperature observed in% INS experiment. The solid line is 
%a fit using again the  solution of (a).\\

\end{document}